\documentclass[letterpaper]{sfchem}
\usepackage{graphicx}

\newcommand{\ee}[1]{\mbox{${} \times 10^{#1}$}}
\newcommand{\eten}[1]{\mbox{$10^{#1}$}}

\newcommand{\kms}{\mbox{km s$^{-1}$}}
\newcommand\cmv{\mbox{cm$^{-3}$}}

\newcommand{\micron}{$\mu$m}

\newcommand{\hst}{\mbox{\it HST}}
\newcommand{\mm}{millimeter}
\newcommand\submm{submillimeter}
\newcommand\smm{submillimeter}

\newcommand{\mir}{mid-infrared}
\newcommand{\nir}{near-infrared}

\newcommand{\msun}{\mbox{M$_\odot$}}

\newcommand{\tex}{\mbox{$T_{\rm ex}$}}

\newcommand{\tk}{\mbox{$T_K$}}
\newcommand{\td}{\mbox{$T_d$}}

\newcommand{\mean}[1]{\mbox{$\langle#1\rangle$}} 

\newcommand{\form}{H$_2$CO}

\newcommand{\cooo}{C$^{18}$O}

\newcommand{\hcop}{HCO$^+$}
\newcommand{\hcopi}{H$^{13}$CO$^+$}

\newcommand{\jj}[2]{\mbox{$J = #1\rightarrow#2$}}

\input{epsf}

\newcommand{\Tdr}{\mbox{$T_{d}(r)$}}
\newcommand{\Tkr}{\mbox{$T_{K}(r)$}}

\newcommand{\vinf}{\mbox{$v_{inf}$}}
\newcommand{\vinfr}{\mbox{$v_{inf}(r)$}}
\newcommand{\rinf}{\mbox{$r_{inf}$}}
\newcommand{\vturb}{\mbox{$\Delta v_{turb}$}}
\newcommand{\vrot}{\mbox{$v_{rot}$}}
\newcommand{\mstar}{\mbox{$M_{\star}$}}
\newcommand{\apj}{\mbox{ApJ}}
\newcommand{\apjs}{\mbox{ApJS}}
\newcommand{\apjl}{\mbox{ApJL}}
\newcommand{\aj}{\mbox{AJ}}
\newcommand{\aap}{\mbox{Astr. Ap.}}
\newcommand{\mnras}{\mbox{MNRAS}}
\newcommand{\araa}{\mbox{ARAA}}

\begin{document}


\title {Studying Infall}
\author {Neal J. Evans II\inst{}}
\institute{The University of Texas at Austin, Astronomy Department, 
       1 University Station C1400  Austin, Texas 78712--0259}

\authorrunning{Evans }
\titlerunning{Studying Infall}

\maketitle
 
\begin{abstract}

The study of protostellar collapse or infall has proven
unusually difficult and controversial. Some historical context
will be provided, against which recent advances can be measured.
We now have a substantial number of objects with signatures that
can be interpreted in terms of collapse, but a number of issues
remain. One issue is the effect of chemical variations, especially
depletion in the dense, cold interiors of cores that are likely to
form low mass stars. Strategies for dealing with this issue depend
on using dust emission to constrain the density and temperature
distribution, leaving molecular line observations to constrain
dynamics and abundance distributions. Recent progress in this area
will be described and we will consider the next challenges to be
overcome. Interferometeric observations, especially with future
instruments, will provide a powerful tool. In combination with
chemical studies coupled with dynamical models, the observations
made possible by interferometers should finally put this subject
on a firm foundation.


\end{abstract}

\section{Introduction}

While gravitational collapse is the accepted explanation for star formation,
the observational evidence for it is very scarce. Establishing this picture
{\it observationally} is a strong reason for studying infall. This subject
has long been controversial (cf. Goldreich \& Kwan 1974 and Zuckerman \&
Evans 1974), and the burden of proof has rested heavily on those who 
interpret observations in terms of collapse. While it will always be
difficult to rule out other explanations for observational signatures of
collapse, the collapse interpretation is beginning to benefit 
from the principle of simplicity: other explanations are becoming 
increasingly contrived.

If we can
establish the basic picture, we can move on to testing particular theories
for how the collapse proceeds. There are many theoretical models
that differ in their predictions of the density and velocity fields.
In principle, these predictions may be subject to observational tests.
As an example, collapse according to the Larson-Penston similarity
solution (Larson 1969; Penston 1969) produces much wider linewidths 
than inside-out collapse from a singular isothermal sphere (Shu 1977).
The line profiles expected for the Larson-Penston solution do not match
data from regions forming low mass stars (Zhou 1992) but may find application
in regions forming massive stars.

Finally, observational constraints on the dynamics will provide an
alternative way to assess the timescales for star formation. At present,
these are determined only by statistical arguments based on the number
of objects in different observational classes, assumed to be an evolutionary
sequence. This method is subject to a number of uncertainties, including
selection biases and the possible importance of factors other than age, 
such as environment, etc. For example, the usual
view is that protostars spend $2\ee4$ yr in the Class 0 phase and 
$2\ee5$ yr in the Class I phase. The former number is based on the
fact that the $\rho$ Ophiuchus cloud has ten times as many Class I objects
as Class 0 objects (Andr{\'e} \& Montmerle 1994). 
However, Visser, Richer, \& Chandler (2001, 2002)
have found as many Class 0 objects  
as Class I objects in a recent survey of other regions, implying a timescale
for the Class 0 stage that is 10 times longer than usually assumed. 
Direct observation of
infall speeds in various objects would provide an independent estimate
of timescales.

The topic of infall has been reviewed by Zhou \& Evans (1994), Myers (1997),
Zhou (1999), Evans (1999), and Myers, Evans, \& Ohashi (2000). 
In this paper, I intend mostly to update those reviews, but I will begin 
with some background.

\section{Dark Ages and Renaissance}

Partly because of the historical controversy referred to above, the
study of collapse suffered through the ``dark ages" from the mid-1970s
to the mid-1990s. During this period, every claim to find evidence
for collapse was followed by counter-claims with alternative explanations
[cf. Goldreich \& Kwan (1974), Liszt et al. (1974) vs. Zuckerman \& Evans
(1974); Snell \& Loren (1977) vs. Leung \& Brown (1977); Walker et al. (1986)
vs. Mundy, Myers, \& Wilking (1986) and Menten et al. (1987)]. 
For further historical discussion and references, see Evans (1991) 
and Myers et al. (2000). The net effect of all this contention was to make
the study of infall seem disreputable.

In addition to sociological factors, physical factors have
made the field difficult. Collapse motions tend to be small compared to
other motions and hence easily obscured. A rough estimate of the
infall speed is given by the following:

\begin{equation}
\vinf \sim 1\ \kms \Bigl[\frac {\mstar / \msun}{r/1000 {\rm AU}}\Bigr]^{0.5}  .
\end{equation}

Thus, for low-mass stars, one needs to probe small radii to see
substantial infall motion. Other motions are often substantially larger.
For a sample of low-mass cores, Caselli \& Myers (1995) found that the
non-thermal (turbulent) contribution to the linewidth (FWHM) could be fitted by
the following relation:

\begin{equation}
\vturb = 1.3\ \kms r_{pc}^{0.5} .
\end{equation}

While this relation does not necessarily apply to the linewidth as a function
of radius for a {\it given} region (see Goodman et al. 1998), there is some
evidence that turbulent linewidths are less at small radii on average.
Thus improved resolution is an obvious route to improve the detectability
of infall. However, the decreases in linewidth will be limited as thermal
broadening takes over.  The minimum possible linewidth (aside from narrowing
by maser emission) is

\begin{equation}
\Delta v_{th} = \sqrt{\frac{8 ({\rm ln}2) k\tk}{m}} = 
0.215\ \kms \sqrt{\frac {\tk}{m({\rm amu})}} ,
\end{equation}
where \tk\ is the gas kinetic temperature and $m$ is the mass of the species
whose line is being observed. Based on this equation, one would gain by
observing heavier molecules, but purely thermal broadening is not usually
seen in heavier molecules. In fact, turbulence often persists at the
subsonic level even on small scales. The sonic linewidth is

\begin{equation}
\Delta v_{sonic} = \sqrt{\frac{8 ({\rm ln}2) k\tk}{m_{mean}}} = 
0.14\ \kms \sqrt{\tk} ,
\end{equation}
where $m_{mean}$ is the mean mass per particle. In some very quiescent
clouds, the turbulence is considerably less than this value.

In addition, rotational motions will increase as regions closer to the
forming star are probed. If angular momentum is conserved,

\begin{equation}
\vrot \propto r^{-1} .
\end{equation}
Ohashi et al. (1997) found that angular momentum appears to be conserved
inside about 6000 AU. On much smaller scales, where the circumstellar
disk forms, angular momentum will be redistributed, but 
the rotational velocity does appear to increase with decreasing radius
in the infall region.

Finally, and worst, outflows seem to begin very soon after a central source
has formed, and they are very energetic at early times. These move matter
at velocities ranging up to a few times 100 \kms\ in the wind and a few 
times 10 \kms\
in the molecular outflow. Most troubling for the study of infall is the
{\it low} velocity motion of dense gas very near the base of the outflow. These
motions can easily obscure or confuse infall signatures.

The renaissance in the study of infall can be traced to four developments.
First, the discovery of objects in very early stages provided more fertile 
grounds for infall studies. Presaged by extensive surveys for dense gas in
nearby dark clouds by Myers and colleagues (e.g., Myers, Linke \& Benson 1983), the identification of
Class 0 objects by Andr{\'e} et al. (1993) and Class $-1$ objects
(also called Pre-Protostellar Cores or PPCs) by Ward-Thompson et al. (1994)
was a crucial development.
Second, simple models for the density and velocity fields during collapse were
presented by Shu (1977), but the idea of inside-out collapse from an
initial, singular isothermal sphere really came to prominence after the
seminal review article by Shu, Adams, \&Lizano (1987).
Third, Zhou (1992) converted the theoretical models of both the inside-out 
collapse
and the Larson-Penston solution into simulated line profiles as a function
of time, clarifying the characteristic line shapes expected for collapse.
Finally, a credible example of collapse was found in the small globule
B335 (Zhou et al. 1993).

\begin{figure}[ht]
\resizebox{\hsize}{!}{\includegraphics[30,320][500,761]{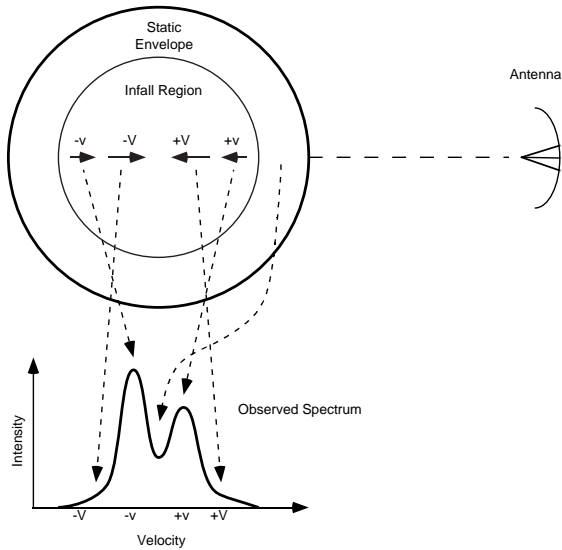}}
\caption{The origin of various parts of the line profile for a cloud
undergoing inside-out collapse. The static envelope outside \rinf\
produces the central self-absorption dip, the blue peak comes from the
back of the cloud, and the red peak from the front of the cloud.
The faster collapse near the center produces line wings, but these are
usually confused by outflow wings.
\label{infallcartoon}}
\end{figure}

\begin{figure}[ht]
\rotatebox{90}{\resizebox{\hsize}{!}{\includegraphics[50,50][550,650]{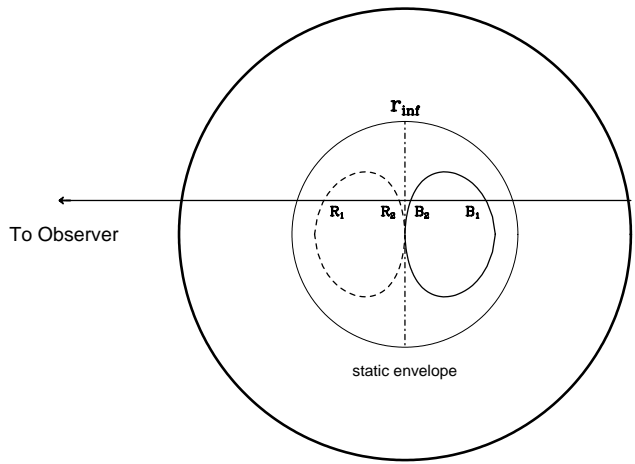}}}
\caption{A schematic explanation of why line profiles of optically thick,
high-excitation lines are skewed to the blue in a collapsing cloud.
The ovals are loci of constant {\it line-of-sight} velocity, for
$v(r) \propto r^{-0.5}$. Each line of sight intersects these loci at
two points. The point closer to the center will have a higher \tex,
especially in lines that are hard to excite, so that
$\tex(R_2) > \tex(R_1)$ and $\tex(B_2) > \tex(B_1)$. If the line is
sufficiently
opaque, the point $R_1$ will obscure the brighter $R_2$, but $B_2$ lies
in front of $B_1$. The result is a profile with the blue peak stronger
than the red peak (Zhou \& Evans 1994).
\label{vloci}}
\end{figure}

The key signature of infall is a line profile that ideally is double-peaked
with a central self-absorption dip, with the blue peak being stronger than
the red peak (Leung \& Brown 1977, Zhou 1992, Myers et al. 1996). 
Cartoons that demonstrate why this signature arises can
be found in Figs. \ref{infallcartoon} and \ref{vloci}. This signature will
appear only if the line has a suitable optical depth and critical density.
Observations of an optically thin line that peaks near the velocity of the
self-absorption dip are necessary to rule out two separate velocity components.
Some optically thick lines will not show a self-absorption dip, 
but will show a blue-skewed profile. The latter are referred to by
Myers et al. (1996) as ``red-shoulder" profiles, but they are the same 
as the ``blue-skewed" profiles.
For simplicity, I will refer to either of these as ``blue profiles."

The full double-peaked signature was seen clearly in several lines toward
B335 by Zhou et al. (1993) and modeled in more detail by
Choi et al. (1995).  While some issues have arisen later regarding B335 (\S 4),
it was credible enough to return some respectability to the field, and it
led to the next development.

\section{Surveys for Infall Signatures}

One example is insufficient to verify a theory for low-mass star formation,
so a number of surveys began for blue profiles.
These initially used the molecular tracers that showed blue profiles
in B335, CS and \form. Various lines were used for the optically thin tracer.
In a survey of 12 globules, similar to B335, Wang et al. (1995) found 3
blue profiles using \form\ as the optically thick tracer. 
Mardones et al. (1997) surveyed a large number of Class 0 and I sources,
finding 14 out of 37 blue profiles in CS and 15 out of 47 in \form.
Most surprisingly, Lee et al. (1999) found blue profiles in 
17/70 starless cores, using CS. In the inside-out picture, a central
luminous object forms almost as soon as collapse begins, so a phase with
collapse and no central luminosity would be very rare.

\begin{figure}[ht]
\resizebox{\hsize}{!}{\includegraphics{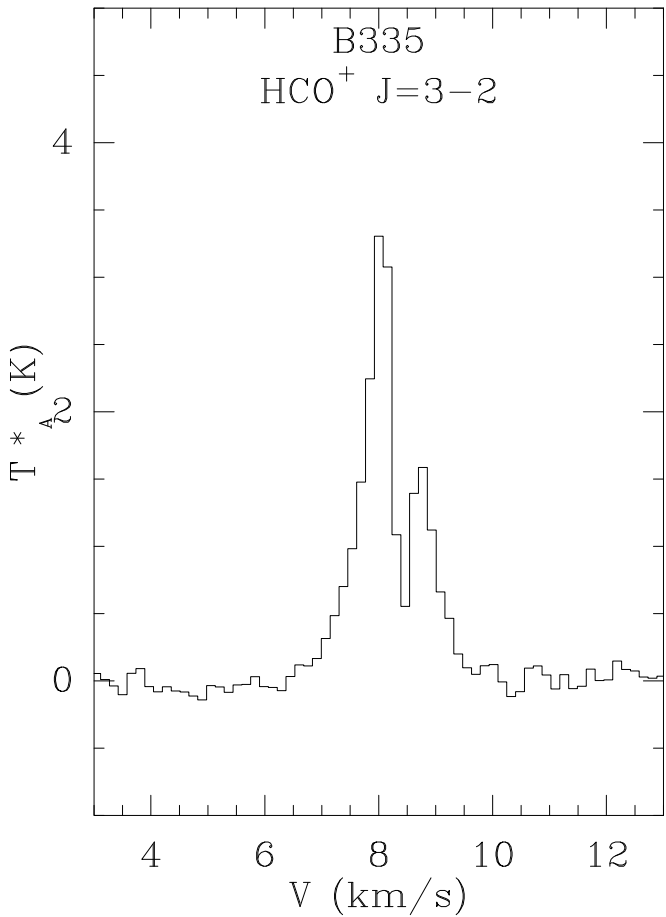}}
\caption{The line profile of the \jj32\ \hcop\ line toward B335 (Evans,
Rawlings, \& Choi 2002b).
\label{b335hcop}}
\end{figure}

Meanwhile, it turned out that \hcop\ showed even deeper
self reversals (Fig. \ref{b335hcop})
 than CS and \form\ toward B335, and searches commenced using
\hcop\ \jj32 as the optically thick line and \hcopi\ as the optically thin
line. Gregersen et al. (1997) found blue profiles toward 6 out of 18 
Class 0 sources.  Because the study by Mardones et al. (1997) 
using CS and \form\ had found fewer blue profiles among the Class I sources
than among Class 0 sources, it was important to put
the Class 0/I searches on a common footing. A study of Class I sources
by Gregersen et al. (2000) found blue profiles in 8 out of 16 Class I
sources and Gregersen \& Evans (2000, 2001) found blue profiles in 6 of 17
Class $-1$ cores (starless cores with \smm\ emission). Clearly the phenomenon
of blue profiles was widespread, but far from universal. Worse yet, some
lines showed {\bf red} profiles!

To quantify the results, Mardones et al. (1997) introduced the excess:
$E = (N_{blue}-N_{red})/N_{tot}$, where $N_{blue}$ and $N_{red}$ are the
number of blue and red profiles in the total sample of $N_{tot}$ objects.
A line was counted as blue if the peak of the optically thick line was
shifted blueward of the peak of the optically thin line by a quarter of the
linewidth of the optically thin line:
$\delta v = (v_{thick}-v_{thin})/\Delta v_{thin} < -0.25$.
Similarly, a red line would have $\delta v > 0.25$.
The excess should be zero on average if the skewing of line profiles were
caused by random outflow blobs or other effects without some overall
inward motion.

The surveys did reveal positive excesses, suggesting a systematic tendency
toward inward motions. One question that arose was whether there was any
tendency for the excess to vary with time. In some pictures, the main infall
occurs during the Class 0 stage, with much less infall occurring during
the Class I stage (e.g., Andr{\'e} et al. 2000). Studies with CS and \form\
did find a lower value of $E$ among Class I sources (Mardones et al. 1997), 
but studies with \hcop\ found the same $E = 0.31$ in Class I sources as
in Class 0 sources (Gregersen et al. 2000). 
Because \hcop\ is more opaque than CS or \form, it
may be that it was better able to reveal infall at later stages when less
material remains in the envelope. Using \hcop, a surprisingly large
excess was found for Class $-1$ cores by Gregersen \& Evans (2000), but
the sample was very small. 
The value in Table 1 is based on counting only $\delta v$ that are less
than $-0.25$ by at least one $\sigma$; if all those with $\delta v < -0.25$
are counted, $E = 0.50$.

Another approach taken with surveys is to ask if different tracers agree.
Gregersen et al. (2000) assembled a table of 40 sources, most with data in 
CS, \form, \hcop, and HCN. There were some cases in which the tracers
disagreed, but considerably more in which they agreed. In quite a few cases,
the \hcop\ showed blue profiles, while the others showed no asymmetry.
Gregersen et al. (2000) suggested that the \hcop\ remained optically
thick enough to develop a blue profile at later stages, when the CS and
\form\ were no longer opaque enough. They presented some simple simulations
of line profiles to support this suggestion. A better understanding of
chemical effects in these regions is needed to put this suggestion on
firmer footing.

\begin{table}  
  \begin{center}
    \caption{Excess versus Class.}\vspace{1em}
    \renewcommand{\arraystretch}{1.2}
    \begin{tabular}[h]{lrcc}
      \hline
      Class  & $-1$ & 0    & I \\
      \hline
      Excess & 0.30 & 0.31 & 0.31 \\
      \hline \\  
      \end{tabular}
    \label{excess}
  \end{center}
\end{table}

\section{Complications}

While the surveys indicated a substantial excess of blue profiles over
red profiles, the presence of red profiles raises a red flag. Without
understanding those profiles, how can we be sure what the blue profiles mean?
The red profiles are likely to be caused by dense clumps in outflowing gas.
Clearly some of the blue profiles could be caused by the same thing.
While the statistics suggest that the predominance of blue profiles
indicate collapse, a blue profile in an individual core may not be 
a reliable signature of collapse.

\begin{figure}[ht]
\resizebox{\hsize}{!}{\includegraphics{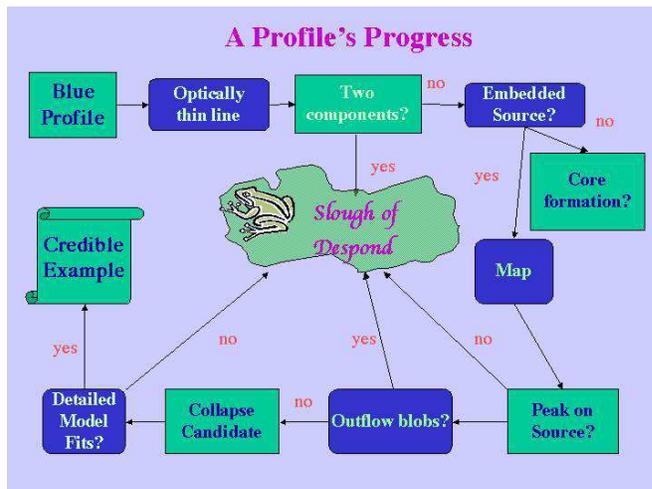}}
\caption{The progress of a blue profile through the many pitfalls on the
path toward ``salvation," as a credible example of collapse (with apologies
to John Bunyan).
\label{progress}}
\end{figure}

In view of the troubled history, modern students of infall have generally
been quite cautious, subjecting a blue profile to a number of tests before
describing it as a credible example of collapse. This process reminds me of
the classic work of Puritan allegory (Bunyan 1688), so I offer a modern
allegory for the ``Profile's Progress" in Fig. \ref{progress}. 
A blue profile could just represent the presence of two velocity
components, which must be ruled out by observations of an optically
thin line that peaks near the dip in the optically thick line, rather than
showing two peaks. If there is an embedded source (Class 0 or I), a map
of the source should reveal a peak on or near the source; the peak could
be in the ``blueness," measured in various ways, or in the depth of the dip. 
The peak of the blue component need not be on the source if rotation is 
also present, but a pattern consistent with expectations for rotation
should be seen. Now, the hardest test appears; a blue
profile must resist the temptation to just be an outflow blob! The profile
that passes this test may be considered a candidate for collapse. Agreement
with someone's detailed model is required to achieve the status of a credible
example. Opinions differ as to whether any regions have achieved this status.

If no embedded source is known (Class $-1$), the path is slightly different
(no outflow to worry about) and is not shown in Figure \ref{progress}. 
First, without a source, one has to check instead that some indicator
is at least centered on the peak of the \submm\ continuum emission.
In these cases, one may be seeing core formation, rather
than star formation, and the pattern of blue profiles may be quite extended
(e.g., Tafalla et al. 1998).  The blue profiles cannot be tracing infall 
onto a central object, so they must be
compared to models of core formation by ambipolar diffusion or turbulent
dissipation (e.g., Myers \& Lazarian 1998) or to models of collapse
before a central object forms (Lee et al. 2003).

Because infall velocities should increase at smaller radii, it seemed that
interferometers might provide clearer evidence of infall. Indeed, some
studies did look promising (Chandler \& Sargent 1993, Zhou et al. 1996, 
Wilner et al. 1997). 
However, Choi et al. (1999) found some discrepancies. In particular, the
red peak was very weak or even absent in some cases. This was an extreme
example of a trend already seen in single-dish data toward higher 
blue/red peak ratios in the data than predicted by the models (Gregersen et al.
1997). In addition, interferometers can artificially produce what
appear to be self-absorption dips by resolved out the extended emission 
from a static or slowly moving envelope (e.g., Gueth \& Guilloteau 1999).

When the CS \jj54 line was imaged in B335 by Wilner et al. (2000), they
found that the line profile, while blue, was narrower than predicted
from the best fitting model to the single dish data (Choi et al. 1995).
At a minimum, this result suggested lower velocities at small radii
than predicted from the inside-out collapse model. Even worse, the
map showed that the blue emission was displaced from the infrared source
and thus more likely an outflow blob.

Around the same time, it was becoming clear that many molecules can be
heavily depleted in the dense, cold regions of Class $-1$ sources (e.g., 
Caselli et al. 1999), and
that these depletions might affect the line profiles strongly (e.g., Rawlings
and Yates 2001) if they persist into the infalling stage. 

The detection of blue profiles in many Class $-1$ cores also called into
question the basic model of inside-out collapse, suggesting instead that
inward motions of some kind are commonly seen before formation of a central
luminosity source. Maps of the lines indicated extended regions of blue
profiles, inconsistent with very early stages of an inside-out collapse.
(Tafalla et al. 1998; Lee, Myers, \& Tafalla 1999, 2001)

All these developments suggested the need for a more robust tracer
of the density structure than provided by molecular lines because
detailed simulation of the line profiles required models of the density,
temperature, velocity, and abundance structure.

\section{Using Dust and Gas}

While dust emission or absorption provides no direct information on velocities,
studies of dust can be quite useful for the infall problem.
Studies of dust continuum emission or absorption 
can constrain the distribution of density and temperature,
cutting down the number of free parameters in models of line profiles.
Conveniently, several new instruments became available in the late 1990s
that revolutionized observations of dust emission. For low-mass, cold
regions, SCUBA on the JCMT, operating at 850 and 450 \micron, and
various bolometer arrays on the IRAM 30-m dish have provided many maps of
\submm\ or \mm\ emission
(e.g., Shirley et al. 2000; Ward-Thompson et al. 1999; Visser et al. 
2001, 2002).

Studies of cores through absorption (or scattering) by dust are complementary 
to emission studies.
These have a long heritage, but one has to use
absorption in the infrared to probe the denser regions. Studies in the
\nir\ by Lada et al. (1994) and in the \mir\ using ISO (Bacmann et al. 2000)
are examples of this technique. The use of the \mir\ will be extended by
observations with SIRTF of dense cores (Evans et al. 2002a).

Proper interpretation of the emission data requires knowledge of the dust
temperature distribution (\Tdr) because at the low temperatures around
low luminosity sources, the dust emission is quite sensitive to temperature.
Unlike the case of luminous sources, the emission is not in the Rayleigh-Jeans
limit and the \Tdr\ is not a power law (Shirley et al. 2000); 
consequently, convenient analytical approaches are inappropriate.  
Fortunately, there are publicly available codes to compute \Tdr\ 
self-consistently for a given density distribution, 
including both internal and external heating (Egan, Leung, \& Spagna 1988,
Nenkova, Ivezi{\'c}, \& Elitzur 2000).

For line profile modeling, one needs to know \Tkr, the distribution
of gas kinetic temperature. For much of the relevant range of radii,
the density is high enough that collisions of molecules with dust grains
makes $\tk \sim \td$, but in the lower density envelope, a separate
energetics calculation is needed to obtain the kinetic temperature. While
one might worry that heavy depletion of molecular coolants deep in the
interiors might affect \tk, Goldsmith (2001) has shown that the effects
are small. Once the density is high enough to affect the coolants significantly,
it is high enough to enforce $\tk \sim \td$ to reasonable accuracy.

The results of models with self-consistent calculations of \Tdr\ (Leung 1975,
Evans et al. 2001) show that Class $-1$ cores are very cold in the center
($\td \sim 7$ K) and warmer on the outside ($\td \sim 14$ K if exposed to
a typical interstellar radiation field). Density distributions that are 
constant in the center but approach power laws at large radii, such as
Bonnor-Ebert spheres fit the radial intensity profile well (Ward-Thompson,
et al. 1999, Evans et al. 2001). The low central \td\ decreases
the emission from the center, even at \mm\ and \smm\ wavelengths.
When the lower temperatures in the center are accounted for, 
the central densities of Class $-1$ cores with reasonably strong \smm\ emission
range from 3\ee5 to \eten6 \cmv\ (Evans et al. 2001). 

At the very low temperatures and high densities deep in Class $-1$ cores, 
substantial depletion of the molecules onto dust
grains is expected on relatively short timescales. The observational 
evidence clearly confirms this expectation (e.g., Rawlings 2000, Caselli 2002 
for reviews). 
Many Class $-1$ cores are nearly
invisible in the usual molecular tracers of column density, such as \cooo.
Other species that form from less condensible precursors or that profit
from the depletion of their primary destroyers follow the dust emission
more closely (e.g., Lee et al. 2003). Especially favored are molecules derived
from N$_2$, ions, and deuterated species. On these grounds, N$_2$D$^+$
may be the last gaseous molecular tracer still working after the others have 
chilled out (Caselli et al. 2002). 

This depletion makes it very hard to trace the
velocity field in the interior, but studies at larger radii show widespread
regions of blue profiles, as described earlier. These profiles can be
successfully modeled in some cases with a Plummer-like model (Whitworth
\& Ward-Thompson 2001); in particular, Lee et al. (2003) have fit the
\hcop\ line profiles toward L1544 at several positions with such a model.
The density and temperature distributions of this model are nearly
identical to the best fitting Bonnor-Ebert sphere and thus predict dust
emission that matches observations, but the Plummer-like model also provides
a velocity distribution needed to compare to line profiles.
In the case that fits L1544, the inward velocity peaks around 0.015 pc.
An alternative approach (Bourke et al. 2002) to the same core
treats each molecular transition with a two-layer model; when applied
to different molecular transitions, one infers a velocity field larger
at smaller radii, qualitatively consistent with the modeling by Lee et al. 
(2003). The exciting thing about these results is that we are beginning
to probe velocity fields in very early stages. These will set timescales
and initial conditions for the formation of central objects. However, these
are only initial steps and much work remains, even to show that these motions
are necessarily related to collapse.

Once a central object forms, the core is heated from the inside as well
as the outside, and \td\ rises in the core. The \smm\ emission from dust
becomes stronger and easier to map. Analysis of a substantial sample
of Class 0 and I objects (Shirley et al. 2002; Young et al. 2003)
finds that the radial intensity profile of the \smm\ emission is 
fitted well by power laws in density [$n(r) = n_f(r/r_f)^{-p}$].
An interesting dichotomy appears: for sources that are nearly spherical,
$\mean p = 1.8$; for the small number of sources that are quite elongated, 
typical $p \sim 1$. While the mean value of $p$ is similar to that of
the singular isothermal sphere, these objects should have evolved into 
the inside-out collapse stage, which predicts $p \leq 1.5$ inside \rinf. 
No sign of a shift to a smaller value of $p$ at small radii is apparent 
in the data from single dishes, which probe $r \geq 2000$ AU, in Class 0
sources (Shirley et al. 2002). More surprisingly, most Class I sources
also show no sign of a change in $p$ outside 2000 AU (Young et al. 2003), even
though \rinf\ should be larger in these sources.

For B335, the inside-out collapse model that fits the molecular line 
data (Choi et al. 1995) does not fit the dust continuum data, which prefer
a power law density. Studies of extinction in the \nir\ using \hst\ 
observations (Harvey et al. 2001) 
find a power law with $p = 2$ in the outer parts of B335,
with some evidence for a turnover near the \rinf\ found by Choi et al. (1995).
The extinction method runs out of stars inside $r \sim 3500$ AU, but recent
studies of dust emission with the IRAM interferometer probing scales from
500 to 5000 AU were fitted well by a power law with $p = 1.65\pm 0.05$
(Harvey et al. 2002).
These results suggest that the inside-out model does reproduce the
density distribution, returning the issue to the models of molecular lines.

Combining the constraints from dust continuum emission with those of
molecular lines in a single, coherent model is clearly necessary.
Hogerheijde \& Sandell (2000) demonstrated the power of this technique; 
they found good fits to models of inside-out collapse for three of four 
sources. The fourth source, L1489 IRS, was later modeled as a more evolved
object with a rotating, contracting disk (Hogerheijde 2001). Such studies
are clearly necessary to reduce the large number of free parameters once
molecular depletion is considered.

\section{The Role of Interferometers}

While data from interferometers must be interpreted with care (see discussion 
in \S 4), they provide the only probe of the velocity fields on small scales.
Choi (2002) has explored some of the systematic effects and simulated line
profiles from infall as seen by interferometers such as the SMA.
As interferometers acquire more antennae, their imaging fidelity will improve,
but it will remain important to account for extended emission. 

In addition to looking for infall signatures in optically thick lines, 
interferometers have been used to probe velocity fields in optically thinner
tracers (e.g., Ohashi et al. 1996, Ohashi 1997, Momose et al. 1998).
The signature of infall is a velocity gradient along the projected {\it minor}
axis of a flattened structure that is larger than the disk. Rotation produces
a velocity gradient along the projected {\it major} axis, allowing one to
separate these two motions. Typically, both infall and rotation are 
inferred from the data. As always, confusion with the outflow is
a problem, but this technique is likely to prove powerful as we begin
to understand which species do {\it not} trace the outflow.

In addition to studies of emission, interferometers provide a new possibility.
Because they are sensitive to emission from disks, the continuum emission 
from the disk can serve as a background lamp and, with even higher
spatial resolution and sensitivity, an occulting disk for the back half
of the cloud (Evans 2001). For disk emission that is sufficiently
bright, the front of the cloud will appear in absorption while the rest
of the cloud appears in emission, producing an inverse P-Cygni profile.
With instruments like ALMA, one may be able to resolve the opaque part
of the disk, occulting the back of the cloud altogether, leaving only
the absorption line from the front of the cloud. A redshift in this 
absorption line is the ``smoking gun" for collapse.  Hints of this
absorption were seen toward NGC 1333 IRAS 4A by Choi et al. (1999), but
the IRAM observations of Di Francesco et al. (2001) show a much more dramatic
inverse P-Cygni profile. The remaining question in this source is whether
the absorption arises from the infalling envelope or from a foreground cloud
that just happens to be redshifted. Whatever the outcome in this particular
case, this technique will be routine with instruments like ALMA and will
allow detailed study of the velocity field (Evans 2001).

\section{Studies of the Velocity Field}

For the moment, most studies of the velocity field are still done
with single dishes. By matching line profiles to models over a map
of a source or by modeling the line profiles toward the center for 
lines with different excitation and opacity, it is possible to decode the
velocity field, albeit with ample uncertainties, in part due to unknown
depletion profiles. There are two general approaches to this problem.
One is to take a well-defined theoretical model and try to constrain it
with many observations to determine whether it fits the data with any
reasonable set of input parameters (e.g., Zhou et al. 1993, Choi et al. 1995,
Lee et al. 2003). The other approach is to decode the velocity field
empirically by treating each molecular line in a fairly simple way, but
using the ensemble to constrain \vinfr. 

The latter approach forms the basis for a study of IRAM 04191 by
Belloche et al. (2002); this paper represents the current state of the
art. This source appears to be in a very early stage of infall.
The line profiles of optically thick, hard-to-excite lines 
toward the center show blue profiles, with the self-absorption dip moving
farther redward as the excitation requirements of the line increase.
By excitation requirements, I mean some combination of temperature and
density. Belloche et al. interpret these line profiles in terms of
a constant $\vinf = 0.1$ \kms\ for $r > 3000$ AU, and \vinf\ increasing
inward for $r < 3000$ AU. From line profiles away from the center, they find
evidence of rotation as well, with \vrot\ increasing with decreasing radius
in to $r = 3500$ AU, but constant (with large uncertainties) inside 3500 AU.
This source is destined to become one of the standard test cases for
infall studies, along with B335 and L1544.

\section{Future Prospects}

While many questions remain and skepticism is justified, the field
seems to be moving from arguing about infall to studying it.
The combined constraints of dust and molecular line data will be
important, as will a deeper theoretical understanding of the depletion
of molecules and other chemical effects. Future study of velocity fields,
just getting under way, will require self-consistent modeling tools, especially
2D or 3D radiative transfer. Observations will need to probe a wide range
of spatial scales, and theory will have to provide testable models that
can be generated by observers from publicly available codes to compare
against their data. If we acquire a better understanding of infall
in regions forming low mass stars, we may be in a position to address
the question in the much more complex regions forming massive stars.

\begin{acknowledgements}

I am grateful to M. Choi, T. Bourke, P. Myers, and J. Di Francesco 
for providing early versions of work in progress and for comments.
L. Loinard provided helpful discussions.
I thank the State of Texas, NASA (Grant NAG5-7203), and the NSF
(AST 9988230) for support.

\end{acknowledgements}


\end{document}